\documentclass[a4paper]{spie}

\usepackage{graphicx}
\usepackage{epstopdf}
\usepackage{dcolumn}
\usepackage{bm}

\title{ Theory of plasmonic waves on a chain of metallic nanoparticles in a liquid crystalline host}
\author{Nicholas A. Pike\supit{a} and David Stroud\supit{a}
\skiplinehalf
\supit{a}Department of Physics, The Ohio State University, Columbus, OH 43210 USA
}
\authorinfo{Send correspondence to D. Stroud.  Email: Stroud@physics.ohio-state.edu}

\date{\today}
\begin{document}

\maketitle

\begin{abstract}

A chain of metallic particles, of sufficiently small diameter and spacing, allows linearly polarized plasmonic waves to propagate along the chain. In this paper, we describes how these waves are altered when the liquid crystal host is a nematic or a cholesteric liquid crystal (NLC or CLC) with or without an applied magnetic field. We find that, in general, the liquid crystal host, either NLC or CLC,  alters the dispersion relations of the transverse ($T$) and longitudinal ($L$) waves  significantly from the dispersion relations for an isotropic host.  We show that by altering the director axis of the liquid crystal relative to the long axis of the metallic chain, that the $T$ branch can be split into two non-degenerate linearly polarized branches (NLC host) or two non-degenerate elliptically polarized branches (CLC host).  When an external magnetic field is applied parallel to both the long axis of the metallic particles and the director of the CLC host, we find that the dispersion relations are odd in an exchange in sign for $\omega$ for the non-degenerate elliptically polarized $T$ branches.  That is, the application of an external magnetic field leads to the realization of a one-way waveguide.  

\end{abstract}

\keywords{Surface Plasmons, One-Way Waveguide, Liquid Crystals, Quasi-static Approximation}

\section{Introduction}

Small metal particles have long been known to exhibit resonant excitations known as ``particle'' or ``surface'' plasmons.  Such particles, if they have dimensions small compared to the wavelength of  light, can exhibit sharp optical absorption peaks, which are typically in the near-infrared or the visible.  These absorption peaks play an important role in the optical response of suspensions of metal particles in a dielectric host\cite{maxwell,pelton,maier3,solymar}, and have been the subject of a large amount of study, both experimental and theoretical. 

Because of recent advances in sample preparation, it has become possible to study {\it  ordered} arrays of such  particles\cite{pike13,meltzer,maier2,tang,park}.   For chains of metallic particles, the coupling between the plasmons comes mainly from the electric field produced by the dipole moment of one nanoparticle, which induces dipole moments on the neighboring nanoparticles.  The dispersion relations for both transverse ($T$) and longitudinal ($L$) plasmonic waves can then be calculated in the so-called quasistatic approximation\cite{brong,maier03,park04}, in which the curl of the electric field is neglected.  This approximation is reasonable when the particles are separated by a distance small compared to the wavelength of radiation.   While this approximation neglects some significant coupling between the plasmonic waves and free photons\cite{weber04}, it gives satisfactory results over most of the Brillouin zone. 
 
In this paper, we first review the dispersion relations of propagating plasmonic waves in a liquid crystalline host, as originally calculated by Pike and Stroud \cite{pike13}.  We also outline an extension of this approach to treat the dispersion relations in the presence of an external magnetic field.  For the case of no magnetic field, we briefly show how the dispersion relations for the $T$ and $L$ waves are modified when the metallic chain is immersed in a liquid crystal host.   We consider two types of such hosts: a nematic liquid crystal (NLC) and a cholesteric liquid crystal (CLC).   An NLC is characterized by a uniaxially anisotropic dielectric tensor.  A unit vector parallel to the symmetry axis of this tensor is known as the director.   A CLC can be viewed as a nematic in which the director rotates about an axis perpendicular to the plane (the ``twist axis'') with a characteristic pitch angle.  We note that much of the notation and derivations  used in this paper follow those originally presented in Pike and Stroud\ \cite{pike13}.  Using a simple approximation outlined below, we show that for the NLC case, both the $L$ and $T$ waves are modified when the director of the liquid crystal is either parallel to or perpendicular to the metallic chain.  For the perpendicular case, the two $T$ branches are split into two non-degenerate linearly polarized branches.  For the CLC we show that, while the $T$ branches once again split into two non-degenerate branches, the resulting waves are elliptically polarized, and are described by dispersion relations dependent on the characteristic pitch angle. 

After this brief discussion, we outline the extension of the calculations for the CLC host to include the effects of an external magnetic field.  In particular, we suggest that such a field may lead to a method of creating a one-way waveguide composed of a linear chain of metallic particles. 
Proposals for one-way wave guides are extensive, including proposals for one-way waveguides at the interface between two materials \cite{yu08} and along chains of ellipsoidal particles arranged in a spiral configuration\cite{ Mazor12, Hadad10}.  Our proposal for a one-way waveguide differs from these both in the use of a chain of spherical metallic particles and in the possibility of tuning the dispersion relations via an applied magnetic field.  

The remainder of this paper is organized as follows.  In the next section, we recap the formalism which allows us to approximately calculate the dispersion relations for $L$ and $T$ waves in the presence of  a liquid crystal host.  In Section III, we modify the formalism to include an external magnetic field, in Section IV we provide numerical examples, and we follow these by a brief concluding discussion in Section V. 

\section{Formalism}

\subsection{Nematic Liquid Crystal Host}

We consider a chain of identical metal nanoparticles, each a sphere of radius $a$, arranged in a one-dimensional periodic lattice along the $z$ axis.  The $n^{th}$ particle is assumed centered at $(0, 0, nd)$ ($-\infty < n < + \infty$).  Let us assume that the quasistatic approximation is valid, that is, the electric field is curl free; this case is approximately applicable when both the radius of the particles and the distance between them are small compared the wavelength of light \cite{brong,maier03,park04}.  It is important to note that other corrections to this approximation exists, such as radiative corrections outlined by Weber and Ford\cite{weber04} .

We first consider how the plasmon dispersion relations are modified when the particle chain is immersed in an  anisotropic dielectric, under the influence of an applied magnetic field.    We assume that the host medium  has a dielectric tensor ($\hat{\epsilon}_h$), with principal dielectric constants $\epsilon_\perp$, $\epsilon_\perp$, and $\epsilon_\|$.    The medium inside the metallic particles is assumed to have a dielectric tensor  $\hat {\epsilon}$ with principal diagonal components given by $\epsilon(\omega)$ and off diagonal elements given by $\epsilon_{xy}$.    We assume that $\epsilon(\omega)$ has the drude form:
\begin{equation}
\epsilon(\omega) = 1 - \frac{\omega_p^2}{\omega(\omega+ i/\tau )} \rightarrow 1 - \frac{\omega_p^2}{\omega^2},
\label{eq:epsw}
\end{equation}
 and $\epsilon_{xy}$ has the form
\begin{equation}
\epsilon_{xy}(\omega) \equiv iA(\omega) = -\frac{\omega_p^2\tau}{\omega}\frac{\omega_c \tau}{(1-i \omega \tau)^2} \rightarrow \frac{\omega_p^2 \omega_c}{\omega^3}.
\label{eq:eps_off}
\end{equation}
Here $\omega_p$ is the plasma frequency, $\omega_c = eB/mc$ is the cyclotron frequency, $B$ is the magnitude of the applied magnetic field, $e$ is the charge of an electron, $m$ is the mass of an electron, $c$ is the speed of light, and $\tau$ is a relaxation time, and the last form on the right-hand side of eqs. (\ref{eq:epsw}) and (\ref{eq:eps_off})  applies when $\tau \rightarrow \infty$.  The results below are easily extended to other forms for $\epsilon(\omega)$, as would be required in order to apply this work to nanoparticles composed of real metals such as Au or Ag.

Following the formalism of  Pike and Stroud\ \cite{pike13} we can write down the electric field at ${\bf x}$ due to a sphere with polarization ${\bf P(x^\prime)}$ as \cite{stroud75} 
\begin{equation}
E_i({\bf x}) =  -\int{\cal G}_{ij}({\bf x} - {\bf x}^\prime)P_j({\bf x}^\prime)d^3x^\prime,
\label{eq:polfield}
\end{equation}
where repeated indices are summed over, any external magnetic field is set to zero, and $\cal{G}$ and ${\bf P(x)}$ are defined in Pike and Stroud\ \cite{pike13}. 

To obtain a self-consistent equation for plasmonic waves along a chain immersed in an anisotropic host, we consider the polarization of the $n^{th}$ particle, which we write as ${\bf P}_n({\bf x}) = \hat{\delta\epsilon}\cdot{\bf E}_{in,n}({\bf x})$, where ${\bf E}_{in,n}({\bf x})$ is the electric field within the $n^{th}$ particle.  This field, in turn, is related to the external field acting on the $n^{th}$ particle and arising from the dipole moments of all the other particles.  We approximate this external field as uniform over the volume of the particle,  and denote it ${\bf E}_{ext,n}$.  This approximation should be reasonable if we remain in the region of validity for the quasistatic approximation.   Then ${\bf E}_{in, n}$ and
${\bf E}_{ext,n}$ are related by \cite{Stroud}
\begin{equation}
{\bf E}_{in,n} = (\hat{1} - \hat {\Gamma}\cdot \hat{\delta\epsilon})^{-1}\cdot{\bf E}_{ext,n}.
\label{eq:einext}
\end{equation}
Here $\hat{ \Gamma}$ is a  "depolarization matrix'' defined, for example, in Stroud.\ \cite{stroud75}.   ${\bf E}_{ext,n}$  is the field acting on the $n^{th}$ particle due to the dipoles produced by all the other particles, as given by eq.\ (\ref{eq:polfield}).  Hence, the dipole moment of the $n^{th}$ particle may be written
\begin{equation}
{\bf p}_n = \frac{4\pi}{3}a^3{\bf P}_{in,n} = \frac{4\pi}{3}a^3\hat{ t}\cdot{\bf E}_{ext,n},
\label{eq:pinn}
\end{equation}
where 
\begin{equation}
\hat{ t} = \hat{\delta\epsilon}\left(\hat{ 1}-\hat{ \Gamma}\cdot\hat{\delta\epsilon}\right)^{-1}
\label{eq:tmatrix}
\end{equation}
is a ``t-matrix'' describing the scattering properties of the metallic sphere embedded in the surrounding material.  

Finally, we make the assumption that the portion of ${\bf E}_{ext,n}$ which comes from particle $n^\prime$ is obtained from eq.\ (\ref{eq:polfield}) as if  the spherical particle $n^\prime$ were a point particle located at the center of the sphere within the quasi-static approximation.  With this approximation, and combining eqs.\  (\ref{eq:polfield}),  (\ref{eq:pinn}), and (\ref{eq:tmatrix}), we obtain the following self-consistent equation for coupled dipole moments:
\begin{equation}
{\bf p}_n = -\frac{4\pi a^3}{3}\hat{ t}\sum_{n^\prime \neq n}\hat{{\cal G}}({\bf x}_n - {\bf x}_{n^\prime})\cdot{\bf p}_{n^\prime}.
\label{eq:selfconsist}
\end{equation}

\begin{figure}[ht]
\begin{center}
\includegraphics[scale=1.0]{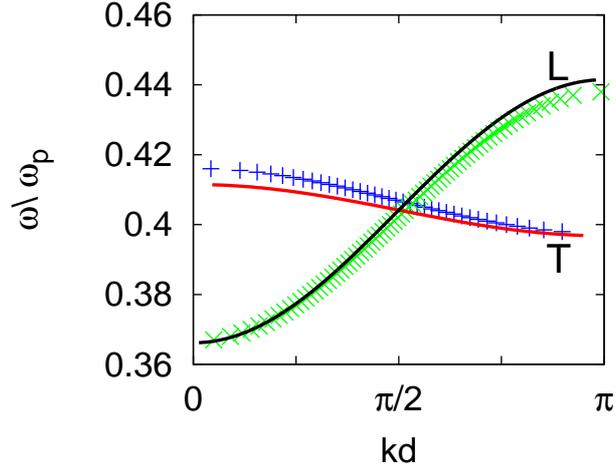}
\caption{(Color Online) Calculated dispersion relations $\omega(k)$ for plasmon waves along a chain of metallic nanoparticles, in the presence of an NLC host.   We plot $\omega/\omega_p$, where
$\omega_p$ is the plasma frequency, as a function of $kd$, where $d$ is the distance between sphere centers.  Green and blue (x's and +'s):  $L$ and $T$ modes for a chain embedded in an NLC with director parallel to the chain.   The NLC is assumed to have principal dielectric tensor elements $\epsilon_\| = 3.0625$ and $\epsilon_\perp = 2.3104$  parallel and perpendicular to the director, corresponding to the material known as E7.   In this and subsequent plots $a/d = 1/3 $, where $a$ is the metallic sphere radius. 
Also shown are the corresponding $L$ and $T$ dispersion relations (black and red solid lines, respectively) when the host is isotropic with dielectric constant $\epsilon_h = 2.5611 = \frac{1}{3}\epsilon_\| + \frac{2}{3}\epsilon_\perp$.} 
\label{figure1}
\end{center}
\end{figure}

Using eq.\ (\ref{eq:selfconsist}), we can determine the dispersion relations for the $L$ and $T$ waves for the case of no external magnetic field, provided the director axis of the liquid crystal coincides with the direction of the metallic chain. In this case, the dispersion relations for the $L$ waves for nearest neighbor interactions between the metallic particles are found to be given by
\begin{equation}
1 = \frac{4}{3}\frac{a^3}{d^3}\frac{\delta\epsilon_{\|}}{1-\Gamma_{\|}\delta\epsilon_{\|}}\frac{1}{\epsilon_{\perp}}\cos kd,
\label{eq:lwave1}
\end{equation}
and for the $T$ waves as
\begin{equation}
1 = -\frac{2}{3}\frac{a^3}{d^3}\frac{\delta\epsilon_{\perp}}{1-\Gamma_{\perp}\delta\epsilon_{\perp}}\frac{\epsilon_{\|}}{\epsilon_{\perp}^2}\cos kd
\label{eq:twave1}
\end{equation}
where the notation of Pike and Stroud  \cite{pike13} has been used.  The calculated dispersion relations given by eqs.\ (\ref{eq:lwave1}) and (\ref{eq:twave1}) are shown in Fig.\ \ref{figure1} for the liquid crystal known as E7, using the principal dielectric constants given by M\"{u}ller \cite{muller}.

As a second example, we also write down the dispersion relations for the case of no external magnetic field when the director of the liquid crystal is perpendicular to the metallic chain.  In this case we get two non-degenerate  linearly polarized $T$  branches and a single $L$ branch.  Again considering the case of nearest neighbor interactions, we find
\begin{eqnarray}
1 & = & -\frac{2a^3}{3d^3}\frac{\delta\epsilon_{xx}}{1-\Gamma_{xx}\delta\epsilon_{xx}}\frac{\epsilon_\perp^{1/2}}{\epsilon_\|^{3/2}}\cos kd, \nonumber \\
1 & = & -\frac{2a^3}{3d^3}\frac{\delta\epsilon_{yy}}{1-\Gamma_{yy}\delta\epsilon_{yy}}\frac{1}{\epsilon_\|^{1/2}\epsilon_\perp^{1/2}}\cos kd,  \nonumber \\
1 & = & \frac{4a^3}{3d^3}\frac{\delta\epsilon_{zz}}{1-\Gamma_{zz}\delta\epsilon_{zz}}\frac{1}{\epsilon_\perp^{1/2}\epsilon_\|^{1/2}} \cos kd,
\label{eq:disp_perp}
\end{eqnarray}
where the notation of Pike and Stroud \ \cite{pike13} is used.  

If we assume that the metallic particle has a Drude dielectric function of the form $\epsilon(\omega) = 1-\omega_p^2/\omega^2$, then the dispersion relation for $L$ and $T$ waves, given in eqs. (\ref{eq:lwave1}), (\ref{eq:twave1}), and those in eq.  (\ref{eq:disp_perp}) neglect damping of the waves due to dissipation within the metallic particles.  To include the effect of damping, one can simply solve eqns.\  (\ref{eq:lwave1}), (\ref{eq:twave1}) or (\ref{eq:disp_perp}) for $k(\omega)$, using the Drude function with a finite $\tau$.   The resulting $k(\omega)$ will be complex in both cases; $[\mathrm{Im}k(\omega)]^{-1}$ gives the exponential decay length of the $L$ or $T$ wave along the chain.  

\begin{figure}[ht]
\begin{center}
\includegraphics[scale=1.0]{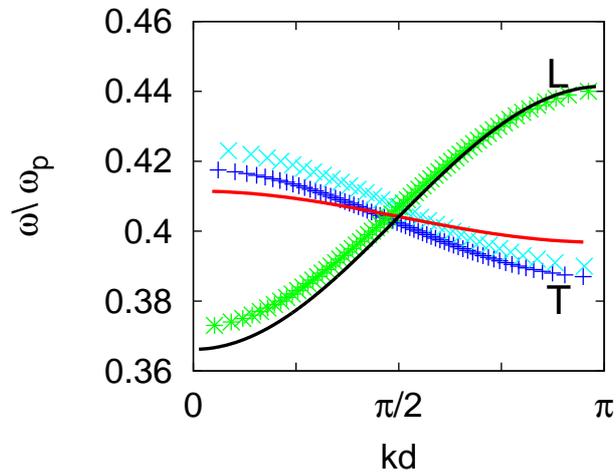}
\caption{(Color Online) Same as Fig.\ 1 except that the director of the NLC is perpendicular to the chain of metal nanoparticles.  The frequencies of
the $L$ modes (asterisks, in green)
and $T$ modes (+'s and x's, shown in dark and light blue), divided by the plasma frequency $\omega_p$,  are plotted versus $kd$.  The NLC has the same dielectric tensor elements as in Fig.\ 1.   Also shown are the corresponding $L$ (solid black) and $T$ (solid red) branches for an isotropic host with $\epsilon_h = 2.5611$.   Note that the $T$ branches which were  degenerate in Fig.\ 1 are split into two branches in this NLC geometry.}
\label{figure2}
\end{center}
\end{figure}

\subsection{Cholesteric Liquid Crystalline Host}\label{chiral_der}

Next, we consider the case where the host is a cholesteric liquid crystal, also known as a chiral nematic liquid crystal.   In such a material, the director, instead of pointing  everywhere in the same direction, rotates in space about an axis perpendicular to the director.   To be definite, we assume that the director ${\bf \hat{n}( x})$ has the following dependence on position:
\begin{equation}
{\bf \hat{n}}({\bf x}) = {\bf \hat{x}}\cos(\alpha z) + {\bf \hat{y}}\sin(\alpha z).
\label{eq:chiral}
\end{equation} 
That is, the director rotates around the $z$ axis with pitch angle $\alpha$.  We still assume that the chain of nanoparticles lies along the $z$ axis.  Thus, the director is always perpendicular to that chain.

As a simple model, we assume that the dielectric tensor of the cholesteric liquid crystal is locally the same as that of an NLC, again assuming the external magnetic field is zero, except that the symmetry axis of the tensor spirals around the $z$ axis with pitch angle $\alpha$.    The tensor can then be written 
\begin{equation}
\hat{\epsilon}(z) = \hat{R}^{-1}(z)\hat{\epsilon} \hat{R}(z),
\end{equation}
where $\hat{\epsilon}$ is the dielectric tensor of a nematic liquid crystal with nonzero components $\epsilon_{xx} = \epsilon_\|$, $\epsilon_{yy} = 
\epsilon_{zz} = \epsilon_\perp$, and $\hat{R}(z)$ is a matrix corresponding to a rotation about the $z$ axis by twist angle $\alpha z$, with non-zero components
\begin{eqnarray}
R_{xx}(z) = R_{yy}(z) = \cos (\alpha z), \nonumber \\
R_{xy}(z) = -R_{yx}(z) = \sin(\alpha z), \nonumber \\
R_{zz}(z) = 1.
\label{rotationmatrix}
\end{eqnarray}
This model is, in fact, a special case of a more general model  given for a cholesteric liquid crystal in Berreman et al.\ \cite{berreman}. 

Next, we calculate the dispersion relations for plasmonic waves propagating along the chain of metallic nanoparticles in the presence of this host, using a simple approximation.   We again start with eq.\ (\ref{eq:selfconsist}).   For simplicity, we include only nearest-neighbor interactions.  Then, for the case of a CLC host,  eq. (\ref{eq:selfconsist}) becomes
\begin{eqnarray}
{\bf p}_n = -\frac{4\pi a^3}{3}[\hat{ t}_{n+1}\hat{{\cal G}}_{n,n+1}\cdot{\bf p}_{n+1} + \nonumber \\
 \hat{ t}_{n-1}\hat{{\cal G}}_{n,n-1}\cdot{\bf p}_{n-1}].
\label{eq:selfconsist_rotate}
\end{eqnarray}
Here $\hat{ t}_{n^\prime}\hat{{\cal G}}_{n,n^\prime}{\bf p}_{n^\prime}$ represents the dipolar field at the position of ${\bf p}_n$ induced by the dipole 
${\bf p}_{n^\prime}$.   To a good approximation, this field will be the same as that found in the case of an NLC host with director perpendicular to the chain, except that both the induced dipole ${\bf p}_{n^\prime}$, and the corresponding dipolar field, are rotated by an amount which increases linearly with position along the $z$ axis.   (This way of expressing the dipolar field is only approximately valid, because the sphere is embedded in a dielectric which is not only anisotropic but also inhomogeneous, varying with position along the $z$ axis.)  With this assumption, the product $\hat {t}_{n^\prime}\hat{{\cal G}}_{n,n^\prime}$ can be written as
\begin{equation}
\hat{ t}_{n^\prime}\hat{{\cal G}}_{n,n^\prime} = \hat{R}^{-1}(z_{n^\prime})(\hat{ t}\hat{{\cal G}}(z_n-z_{n^\prime}))\hat{R}(z_{n^\prime}),
\end{equation}
where $\hat{R}(z)$is given by eq.\ (\ref{rotationmatrix}).

From these equations, we can see that the $L$ and $T$  waves are still decoupled, as in the case of an NLC host with director perpendicular to the chain.   The $L$ modes  are unchanged from the NLC case, but the $T$ modes are altered.  Eq.\  (\ref{eq:selfconsist_rotate}) can be rewritten as
\begin{eqnarray}
{\tilde p}_n = -\frac{4\pi a^3}{3}[\hat{R}^{-1}(z_1)\hat{ t}\hat{\cal G}_{n,n+1}\cdot{\tilde p}_{n+1} + \nonumber \\
\hat{R}(z_1)\hat{ t}\hat{\cal G}_{n,n-1}\cdot{\tilde p}_{n-1}],
\label{eq:selfcon_two}
\end{eqnarray}
where we use $\hat{R}(z_1) = \hat{R}^{-1}(-z_1)$ and ${\bf p}_n = \hat{R}^{-1}(z_n)\cdot \tilde{p}_n$ which corresponds to a rotated dipole moment.

We now write out eq. (\ref{eq:selfcon_two}) explicitly and obtain dispersion relations for the two transverse branches.   We consider only the $x$ and $y$ components of eq.\ (\ref{eq:selfcon_two}), since the $L$ branch is unchanged from the purely NLC case and does not couple to the $T$ branches.   We denote the non-zero diagonal elements of the product matrix $\hat{ t}\hat{{\cal G}}_{n,n+1}$ by $\tau_{xx} = t_{xx}{\cal G}_{xx}$ and $\tau_{yy} = t_{yy}{\cal G}_{yy}$.   

After a little algebra, and using the $2 \times 2$ block of eq.\ (\ref{rotationmatrix}) for $\hat{R}(z_n)$, we obtain the following equation determining the two-component column matrix $ \tilde{p}_0$ whose components are $\tilde{p}_{x0}$, $\tilde{p}_{y0}$:
\begin{equation}
\tilde{p}_0 = -\frac{8\pi a^3}{3}\hat{ M}(k, \omega)\cdot\tilde{p}_0,
\label{eq:disperchiral}
\end{equation}
where $\hat{ M}$ is found to have components
\begin{eqnarray}
M_{xx} & = & \tau_{xx}\cos(\alpha d)\cos(kd), \nonumber \\
M_{xy} & = & -i\tau_{yy}\sin(\alpha d)\sin(kd), \nonumber \\
M_{yx} & = & i\tau_{xx}\sin(\alpha d)\sin(kd), \nonumber \\
M_{yy} & = &  \tau_{yy}\cos(\alpha d)\cos(kd).
\label{eq:comp_nofield}
\end{eqnarray}
Eq.\ (\ref{eq:disperchiral}) has nontrivial solutions if
\begin{equation}
\mathrm{det}\left[\hat{ 1} +\frac{8\pi}{3}a^3\hat{ M}(k,\omega)\right] = 0.
\label{eq:disperchiral1}
\end{equation}  

\begin{figure}[htb]
\begin{center}
\includegraphics[]{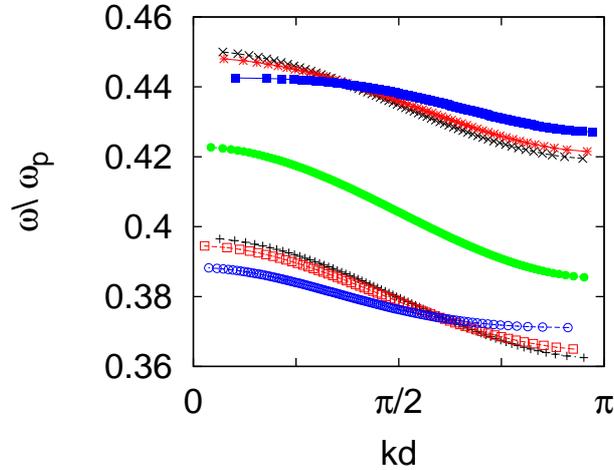}
\caption{(Color Online) Calculated dispersion relations $\omega(k)$ for plasmon waves along a chain of metallic nanoparticles in the presence of a cholesteric liquid crystal host.  We assume that the director rotates about an axis parallel to the chain of metal nanoparticles with a pitch angle $\alpha$.   The red (open square and filled square) plots and blue (crosses and triangles) represent the two $T$ branches for $\alpha d = \pi/6$ and $\pi/3$ respectively, while the black (full circles and asterisks) correspond to $\alpha d = 0$ (nematic liquid crystal). In all cases we assume $\epsilon_\| = 3.06525$ and $\epsilon_\perp = 2.5611$.  The green points (full circles) represent an isotropic host with $\epsilon = \frac{1}{3}\epsilon_\| + \frac{2}{3}\epsilon_\perp = 2.5611$.}
\label{figure3}
\end{center}
\end{figure}

The left-hand side of eq.\ (\ref{eq:disperchiral1}) is  quadratic in $\epsilon(\omega)$ and thus has two solutions for $\epsilon(\omega)$ as a function of $\cos(kd)$.   We plot these two solutions in Fig. \ref{figure3} using values for the dielectric constants from M\"{u}ller\cite{muller} and for various values of the twist angle $\alpha z $. If  $\epsilon(\omega) = 1 - \omega_p^2/\omega^2$,  these lead to two solutions for $\omega$ as a function of $\cos(kd)$, or equivalently, of $k$ (since $\cos(kd)$ is monotonic in k in the range $0<k<\pi/d$).   We use the convention that $\omega > 0$.   If we use the Drude form for $\epsilon(\omega)$ with a finite $\tau$ to include single-grain dissipation, we write $\cos(kd) = [\exp(ikd)+\exp(-ikd)]/2$.   Then there are four solutions for $\exp(ikd)$ and hence for $k$, of which two have $Im(k) > 0$, as required physically.  Thus, this procedure again gives two $T$  branches which include single-particle damping if $\epsilon(\omega)$ has the Drude form with a finite lifetime.

\subsection{Applied Magnetic Field}

We will now discuss how the dispersion relations are changed when there is a CLC host and an external magnetic field is applied parallel to the chain axis, which is also the twist axis of the CLC.   In this case, the dielectric tensor of the metallic host, as given by eq.\ (\ref{eq:eps_off}), is no longer diagonal.  Even though the dielectric tensor of the metallic particle has off-diagonal terms,  the calculation of the dispersion relations again starts from eq.\ (\ref{eq:selfconsist_rotate}) and generally follows the derivation outlined in section \ref{chiral_der}.  

Since the magnetic field is parallel to both the chain axis and the twist axis of the CLC,  the $L$ branch will again decouple from the two $T$ branches.  Furthermore, the $L$ branch will not be modified by the external magnetic field, and behaves similarly to the $L$ branch in the CLC case discussed earlier. The two $T$ branches are expected to be the solutions of a $2\times2$ matrix equation similar to  eq.\  (\ref{eq:comp_nofield}), but including additional terms related to the external magnetic field.   To obtain the dispersion relations in a magnetic field, we need to know how the matrices $\hat{t}_{n+1}$, $\hat{t}_{n-1}$, $\hat{\cal G}_{n,n+1}$, and $\hat{\cal G}_{n,n-1}$ are modified from their zero-field values.   Because the off-diagonal elements of the dielectric tensor satisfy $\epsilon_{ji} = -\epsilon_{ij}$, in a magnetic field, it can be shown that ${\cal G}_{n,n+1}$ and ${\cal G}_{n,n-1}$ are both diagonal and unchanged from their zero-field forms.  We may write these diagonal elements ${\cal G}_{xx}$ and ${\cal G}_{yy}$.   However, $\hat{t}_{n+1}$ and $\hat{t}_{n-1}$ become non-diagonal in a magnetic field and are also independent of $n$, $n-1$, and $n+1$.  We denote this matrix simply $\hat{t}$. We denote the  matrix elements of the $2 \times 2$  projection of $\hat{t}$ by $t_{xx}$, $t_{xy}$, $t_{yx} = -t_{xy}$, and $t_{yy}$.  

After some additional algebra, we find that the non-zero matrix elements of the corresponding $2\times 2$ matrix $\hat{M}( k, \omega)$ 
can be written out explicitly. The matrix $\hat{M}(k,\omega)$ now includes the contribution from the scattering matrix $\hat{t}$, the Green's function $\hat{\mathcal{G}}$, and the elements of the $2 \times 2$ rotation matrix defined in eq. (\ref{rotationmatrix}).   The matrix elements are given as
\begin{eqnarray}
M_{xx} = {\cal G}_{xx}[t_{xx}\cos(kd)\cos(\alpha d) + it_{xy}\sin(kd)\sin(\alpha d)] \nonumber \\
M_{yy} = {\cal G}_{yy}[t_{yy}\cos(kd)\cos(\alpha d) + it_{xy}\sin(kd)\sin(\alpha d)] \nonumber \\
M_{xy} = {\cal G}_{yy}[t_{xy}\cos(kd)\cos(\alpha d) - it_{yy}\sin(kd)\sin(\alpha  d)] \nonumber \\
M_{yx} = {\cal G}_{xx}[it_{xx}\sin(kd)\sin(\alpha d) -t_{xy}\cos(kd)\cos(\alpha d)]. 
\label{eq:magfield}
\end{eqnarray}
With the external magnetic field parallel to the director of the CLC, one can now determine the dispersion relation for the two $T$ waves as nontrival solutions to eq. (\ref{eq:disperchiral1}). 
  
The dispersion relations corresponding to eqs.\ (\ref{eq:magfield}) are again given implicitly by eq. (\ref{eq:disperchiral1}).   We have not, as yet, evaluated these relations numerically, but some qualitative points can already be inferred from the form of the matrix elements.   Of most interest, the presence of a magnetic field will lead to dispersion relations which are {\it non-reciprocal}, i.\ e., $\omega(k) \neq \omega(-k)$ in general.  The magnetic field appears only in the off-diagonal elements $\delta\epsilon_{xy}$ and $\delta\epsilon_{yx}$, which are linear in the field except for very large fields.  From this result, it is found that $t_{xy}$ and $t_{yx}$ have terms which are linear in field, while $t_{xx}$ and $t_{yy}$ are independent of field to that order.  The terms involving $t_{xy}$ and $t_{yx}$ in eq.\ (\ref{eq:magfield}) are multiplied by $\sin(kd)$ and thus change sign when $k$ changes sign. Thus, the secular equation determining $\omega(k)$ is not even in k, implying that the dispersion relations are non-reciprocal. This non-reciprocality disappears when the host is an NLC because then $\alpha = 0$ and the terms involving $\sin(kd)$ vanish.  Although the dispersion relations are non-reciprocal, we do not yet know whether there is a region of one-way propagation, i.\ e., where for certain frequencies waves can only propagate in one direction. In future work, we plan to solve these relations numerically to investigate this intriguing possibility.

\section{Numerical Illustrations}

As a first numerical example, we calculate the plasmon dispersion relations for a chain of spherical Drude metal particles immersed in an NLC and $a/d=1/3$.   We consider two cases: liquid crystal director parallel and perpendicular to the chain axis, which we take as the $z$ axis.   For $\epsilon_\|$ and $\epsilon_\perp$, we take the values found in experiments described in M\"{u}ller\cite{muller}, which were carried out on the NLC known as E7.   For comparison, we also show the corresponding dispersion relations for an isotropic host of dielectric constant which is arbitrarily taken as $\frac{1}{3}\epsilon_\| + \frac{2}{3}\epsilon_\perp = 2.5611$. The results of these calculations are shown in Figs.\ 1  and 2 in the absence of damping ($\tau \rightarrow \infty$ in the Drude expression).   As can be seen, both the $L$ and $T$ dispersion relations are significantly altered when the host is a nematic liquid crystal rather than an isotropic dielectric; in particular, the widths of the $L$ and $T$ bands are changed.   When the director is perpendicular to the chain axis, the two $T$ branches are split when the host is an NLC, whereas they are degenerate for an isotropic host, or an NLC host with director parallel to the chain.   

Next, we turn to the case of a chain of metal particles immersed in a cholesteric liquid cystal.   We assume that the nematic axis lies in the $xy$ plane, the chain runs parallel to the $z$ axis, and the nematic axis rotates about the $z$ axis with a pitch angle $\alpha$.   The calculated dispersion relations for the two $T$ branches are shown in Fig.\ 3 for two different  choices of pitch angle $\alpha$, as well as for $\alpha = 0$, corresponding to the purely nematic case.    For comparison, we also show the dispersion relation for the doubly degenerate $T$ branch when the host is an isotropic dielectric with dielectric constant $\epsilon = \frac{1}{3}\epsilon_\| + \frac{2}{3}\epsilon_\perp = 2.5611$.    We again assume that there is no dissipation.  As is evident, the dispersion relations do depend slightly on $\alpha$.     But while in the case of an NLC host the two $T$ branches are linearly polarized with polarization along the $x$ and $y$ axes, respectively, the two $T$  branches in the cholesteric case are no longer linearly polarized.  Instead, each of these two branches has a polarization vector which rotates as a function of $z$.   We have not attempted to plot this in the figure.

\section{Discussion}

The present  calculations and formalism  leave out several effects which may be at least quantitatively important.    First, in our numerical calculations, but not in the formalism, we have omitted all dipolar couplings beyond the nearest neighbors.   Inclusion of further neighbors will quantitatively alter the dispersion relations in all cases considered, but these effects should not be very large, as is already suggested by the early calculations in Brongersma et al\cite{brong} for an isotropic host.   Another possible effect will appear when $a/d$ is significantly greater than $1/3$, namely, the emergence of quadrupolar and higher quasistatic bands which will mix with the dipolar band and change its shape\cite{park04}.  Even for $a/d >1/3$ the plasmon dispersion relations will still be altered by an NLC or CLC  host in the manner described here.  

The present treatment also omits radiative damping which may be important at certain wave vectors and at long wavelengths\cite{weber04}.  We have not extended the present approach to include such radiative effects but do expect that they will provide only a quantitative change, and not qualitatively change the effects we have described.

From the point of view of applications, the use of a liquid-crystalline host is appealing because liquid crystals are significantly affected by an applied electric field.  For example, the director of an NLC tends to align with the applied electric field.  Thus, the dispersion relations of the propagating plasmonic waves could be, in principle, controlled by such an applied electric field.   This control could be quite useful, for example, in developing filters for propagating plasmonic waves.   The application of an external magnetic field along with an external electric field would allow one to tune the frequency range where only one wave direction is allowed to propagate, this has potential applications in optical relay systems, optical filters, and as a filter for backscattering waves.

To summarize, we have shown that the dispersion relations for plasmonic waves propagating along a chain of closely spaced nanoparticles of Drude metal are strongly affected by external electric and magnetic fields.  When under the influence of only an electric field we find that for both materials, the dispersion relations are affected.  If  the host is a uniaxially anisotropic dielectric (such as an NLC), the dispersion relations of both $L$ and $T$ modes are significantly modified, compared to those of an isotropic host, and if the director axis of the NLC is perpendicular to the chain, the two degenerate transverse branches are split.   If the host is a CLC, with rotation axis parallel to the chain, we find that the $T$ modes are again split into nondegenerate bands, whose dispersion relations now depend on the pitch angle $\alpha$.   However, in contrast to the NLC case, the two $T$ branches are no longer linearly polarized.     These effects suggest that the propagation of such plasmonic waves can be tuned, by subjecting the liquid crystal to a suitable electric field, so as to change the frequency band  where wave propagation can occur, or the polarization of these waves.   This control may be valuable in developing devices using plasmonic waves in future optical circuit design.  Under the influence of an external magnetic field, we find that for a CLC host the $T$ waves are likely to be non-reciprocal, i.\ e., to have a dispersion relation where $\omega(k) \neq \omega(-k)$.
Possibly there may be a frequency range in which only one direction of wave propagation is allowed.  If this is confirmed by numerical calculation, this type of system would have a possible application as a filter for backscattered waves  in an optical system.

\section*{Acknowledgments}

This work was supported by the Center for Emerging Materials at The Ohio State University, an NSF MRSEC  (Grant No.\ DMR0820414).


\begin{thebibliography}{spiebib}

\bibitem{maxwell} Maxwell, J.\ C.\ "Colours in Metal Glasses, in Metallic Films, and in Metallic Solutions. II,"  {\it Phil.\ Trans.\ R.\ Soc.\ Lond.}\ {\bf 155}, 459-512 (1865). 

\bibitem{pelton} For reviews, see, e.\ g., Pelton, M., Aizpurua, J., and Bryant, G., "Metal-nanoparticle plasmonics," {\it  Laser and Photonic Reviews} {\bf 2}, 136 (2008), or the following two references.

\bibitem{maier3} Maier, S.\ A. {\it Plasmonics: Fundamentals and Applications},  (Springer, New York, 2007).

\bibitem{solymar}  Solymar, L.  and Shamonina, F. {\it Waves in Metamaterials}, (Oxford University Press, Oxford, 2009).

\bibitem{meltzer}Maier,  S.\  A., Brongersma, M.\  L.,  Kik, P.\  G., Meltzer, S., Requicha, A. A. G., Atwater, H. A. "Plasmonics - a route to nanoscale optical devices,"  {\it Adv.\ Mat.}  {\bf 13} 1501 (2001).

\bibitem{maier2} Maier, S.\ A., Kik, P.\ G., Atwater,  H.\ A., Meltzer, S., Harel, E., Koel, B.\ E., Requicha, A.\ A.\ G. "Local detection of electromagnetic energy transport below the diffraction limit in metal nanoparticle plasmon waveguides," {\it  Nature Mater.}   {\bf 2}, 229 (2003).

\bibitem{tang} Tang, Z.\ Y. and Kotov, N.\ A., "One-Dimensional Assemblies of Nanoparticles: Preparation, Properties, and Promise," {\it Adv.\  Mater.} {\bf 17}, 951 (2005).

\bibitem{park}  Park, S.\  Y., Lytton-Jean, A.\ K.\ R., Lee, B.,  Weigand, S.,  Schatz, G.\ C., and  Mirkin, C.\ A., "DNA-programmable nanoparticle crystallization," {\it  Nature }{\bf 451}, 553-556 (2008).

\bibitem{pike13} Pike, N. A.  and Stroud,  D. "Plasmonic waves on a chain of metallic nanoparticles: effects of a liquid-crystalline host." {\it J. Opt. Soc. of Amer.  B} {\bf 30} 1127  (2013).

\bibitem{brong}  Brongersma, M. L., Hartman, J. W.,  and Atwater, H. A., " Electromagnetic energy transfer and switching in nanoparticle chain arrays below the diffraction limit," {\it Phys.\ Rev.\  B} {\bf 62}, R16356 (2000).

\bibitem{maier03} Maier, S. A.,  Kik, P. G.,  and Atwater, H. A., "Optical pulse propagation in metal nanoparticle chain waveguides," {\it  Phys.\ Rev.\  B}{\bf  67}, 205402 (2003).

\bibitem{park04} Park, S. and Stroud, D. , "Surface-plasmon dispersion relations in chains of metallic nanoparticles: An exact quasistatic calculation," {\it Phys.\ Rev.\  B} {\bf 69}, 125418(R) (2004).

\bibitem{weber04}  Weber, W. H.  and Ford, G. W. "Propagation of optical excitations by dipolar interactions in metal nanoparticle chains,"  {\it Phys.\ Rev. B} \ {\bf  70}, 125429 (2004).

\bibitem{yu08} Yu, Z., Veronis, G., Wang, Z., Fan, S. " One-Way Electromagnetic Waveguide Formed at the Interface between a Plasmonic Metal under a Static Magnetic Field and a Photonic Crystal,"  {\it Phys. Rev. Lett.} {\bf 100} 023902 (2008).

\bibitem{Mazor12} Mazor, Y. and Steinberg, B. Z. "Longitudinal chirality, enhanced nonreciprocity, and nanoscale planar one-way plasmonic guiding." {\it Phy. Rev. B} {\bf 86} 045120 (2012).

\bibitem{Hadad10} Hadad, Y. and Steinberg, B. Z. "Magnetized Spiral Chains of Plasmonic Ellipsoids for One-Way  Optical Waveguides." {\it Phy. Rev. Lett.}  {\bf 105} 233904 (2010).  

\bibitem{stroud75}  Stroud, D.  "Generalized effective-medium approach to the conductivity of an inhomogeneous material," {\it  Phys.\ Rev.\  B}  {\bf 12}, 3368 (1975).

\bibitem{Stroud} Stroud, D.  and Pan, F. P. " Effect of isolated inhomogeneities on the galvanomagnetic properties of solids," {\it  Phys.\ Rev.\  B}  {\bf 13} 1434 (1976).

\bibitem{berreman}  Berreman, D. W.  and  Scheffer, T. J. "Bragg Reflection of Light from Single-Domain Cholesteric Liquid-Crystal Films," {\it  Phys. Rev.\ Lett.} {\bf 25}, 577 (1970).

\bibitem{muller} J.\ M\"{u}ller, C.\ S\"{o}nnichsen, H.\ von Poschinger, G.\ von Plessen, T.\
 A.\ Klar, and J.\  Feldmann, "Electrically controlled light scattering with single metal nanoparticles ," {\it Appl.\ Phys.\ Lett.}\ {\bf 81}, 171 (2002).



\end{thebibliography}
\end{document}